\documentclass[superscriptaddress,reprint,twocolumn,aps,floatfix]{revtex4-1}

\usepackage{graphicx}
\usepackage{epsfig}
\usepackage{color}
\usepackage{amssymb}
\usepackage{amsmath}
\usepackage{array}
\usepackage[loose]{units}
\usepackage{hyperref}
\usepackage{cleveref}
\usepackage{braket}
\usepackage[caption=false]{subfig}
\usepackage[english]{babel}
\usepackage{letltxmacro}

\LetLtxMacro{\ORIGselectlanguage}{\selectlanguage}
\makeatletter
\DeclareRobustCommand{\selectlanguage}[1]{%
    \@ifundefined{alias@\string#1}
      {\ORIGselectlanguage{#1}}
      {\begingroup\edef\x{\endgroup
         \noexpand\ORIGselectlanguage{\@nameuse{alias@#1}}}\x}%
}
\newcommand{\definelanguagealias}[2]{%
  \@namedef{alias@#1}{#2}%
}
\makeatother

\definelanguagealias{en}{english}

\newcommand{\unittw}[3]{$\unit[#1]{#2 \textrm{#3}}$}
\newcommand{\unitto}[2]{$\unit[#1]{\textrm{#2}}$}

\newcommand{\unitmo}[2]{\unit[#1]{\textrm{#2}}}
\newcommand{\littleI}{{\scriptstyle\mathcal{I}}}
\newcommand{\UO}{Department of Physics, University of Oregon, Eugene, OR 97403, USA}
\newcommand{\GOE}{Georg-August-Universit\"{a}t G\"{o}ttingen, D-37077 G\"{o}ttingen, Germany}
\newcommand{\NCEM}{National Center for Electron Microscopy, Molecular Foundry, Lawrence Berkeley National Laboratory, Berkeley, CA 94720, USA}
\newcommand{\UMICHC}{Department of Chemical Engineering, University of Michigan, Ann Arbor, MI 48109, USA}
\newcommand{\UMICHMS}{Department of Materials Science and Engineering, University of Michigan, Ann Arbor, MI 48109, USA}

\begin{document}
                            
\begin{abstract}
  Efficient imaging of biomolecules, 2D materials and electromagnetic fields depends on retrieval of the phase of transmitted electrons. We demonstrate a method to measure phase in a scanning transmission electron microscope using a nanofabricated diffraction grating to produce multiple probe beams. The measured phase is more interpretable than phase-contrast scanning transmission electron microscopy techniques without an off-axis reference wave, and the resolution could surpass that of off-axis electron holography. We apply the technique to image nanoparticles, carbon substrates and electric fields. The contrast observed in experiments agrees well with contrast predicted in simulations.
\end{abstract}
\title{Interpretable and efficient contrast in scanning transmission electron microscopy with a diffraction grating beamsplitter}
\author{Tyler R. Harvey}
\affiliation{\UO}
\affiliation{\GOE}
\author{Fehmi S. Yasin}
\affiliation{\UO}
\author{Jordan J. Chess}
\affiliation{\UO}
\author{Jordan S. Pierce}
\affiliation{\UO}
\author{Roberto M. S. dos Reis}
\affiliation{\NCEM}
\author{Vasfi Burak \"{O}zd\"{o}l}
\affiliation{\NCEM}
\author{Peter Ercius}
\affiliation{\NCEM}
\author{Jim Ciston}
\affiliation{\NCEM}
\author{Wenchun Feng}
\affiliation{\UMICHC}
\author{Nicholas A. Kotov}
\affiliation{\UMICHC}
\affiliation{\UMICHMS}
\author{Benjamin J. McMorran}
\email{mcmorran@uoregon.edu}
\affiliation{\UO}
\author{Colin Ophus}
\affiliation{\NCEM}
\date{\today}

\maketitle

\section{Introduction}

The electron microscope offers the opportunity to directly image material structure and physical processes at atomic length scales. Whereas many bulk measurements must be interpreted to infer microscopic structure or processes, in the transmission electron microscope (TEM), one can directly measure the atomic number and positions of atoms and atomic columns, local shifts in atomic transition energies, and electronic and magnetic properties with high precision \cite{suenaga_element-selective_2000,muller_atomic-scale_2008,krivanek_atom-by-atom_2010,lovejoy_single_2012,yang_deciphering_2017}. Scanning transmission electron microscopy (STEM) with a high-angle annular dark field detector (HAADF) has long offered highly interpretable contrast at atomic resolution \cite{pennycook_high-resolution_1990,krivanek_atom-by-atom_2010}. 

However, the electron dose required to produce a good signal-to-noise ratio with HAADF-STEM is high even on high-atomic-number materials, and becomes prohibitive for dose-sensitive low-atomic-number materials that weakly scatter electrons and suffer structural damage quickly \cite{egerton_control_2013}. Efficient imaging depends on measurement of the small phase shifts that an electron acquires upon passing through such a specimen. The most common phase-contrast imaging method employs a small defocus for contrast in high-resolution transmission electron microscopy (HRTEM) \cite{glaeser_invited_2013}. However, efficient phase contrast is also possible in STEM. Now that direct election detectors with a rapid frame rate are available, ``4D STEM'' techniques that utilize one diffraction pattern per scanned probe position are much more feasible. Ptychography and matched illumination and detector interferometry (MIDI-STEM) offer a dose-efficient alternative for interpretable phase contrast in STEM \cite{hoppe_beugung_1969,rodenburg_ptychography_2008,yang_enhanced_2016,yang_simultaneous_2016,rose_phase_1974,ophus_efficient_2016}. These two techniques enable reconstruction of the full optical transfer function--amplitude and phase--of a specimen, and therefore offer more efficient contrast on low-atomic-number materials. However, as ptychography, MIDI-STEM and HRTEM are only sensitive to local phase variations, they effectively high-pass-filter the phase. It is therefore difficult to quantitatively measure thickness or long-range electric and magnetic fields. Center-of-mass and differential phase contrast STEM are sensitive to the derivative of phase and can be used for phase-contrast imaging and, when calibrated properly, electromagnetic field measurement \cite{dekkers_differential_1974,rose_phase_1974,chapman_direct_1978,waddell_linear_1979}, but are similarly only sensitive to local phase variations, and interpretation of contrast is not always straightforward \cite{maclaren_origin_2015,clark_probing_2018}.

\begin{figure}[h]
  \subfloat[]{\label{subfig:schematic:biprism}
  \includegraphics[height=1.5in]{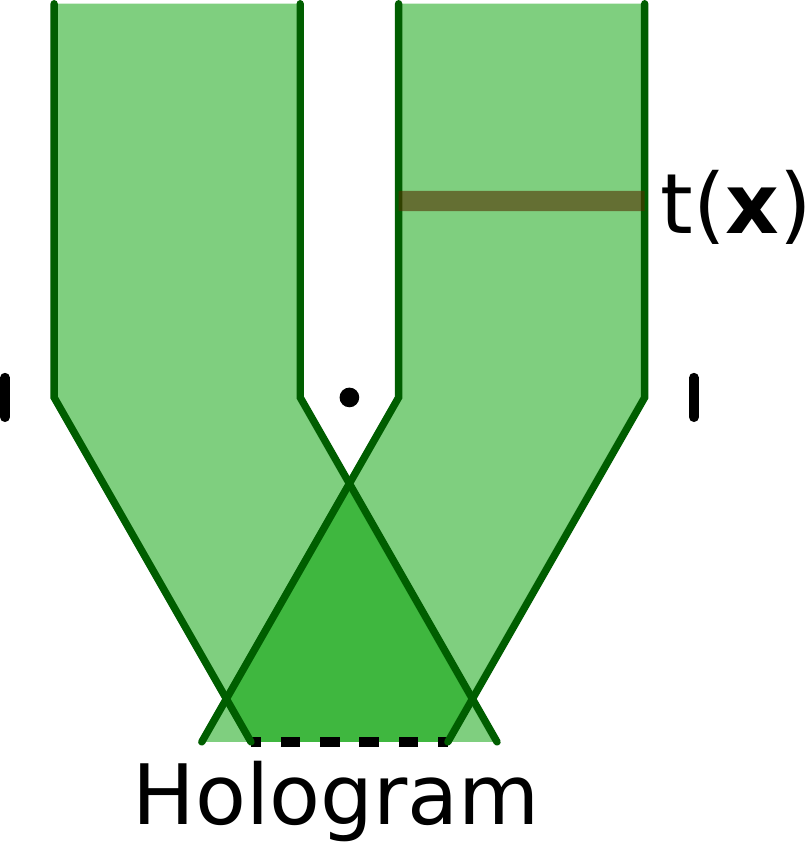}}
  \subfloat[]{\label{subfig:schematic:stemh}
  \includegraphics[height=1.5in]{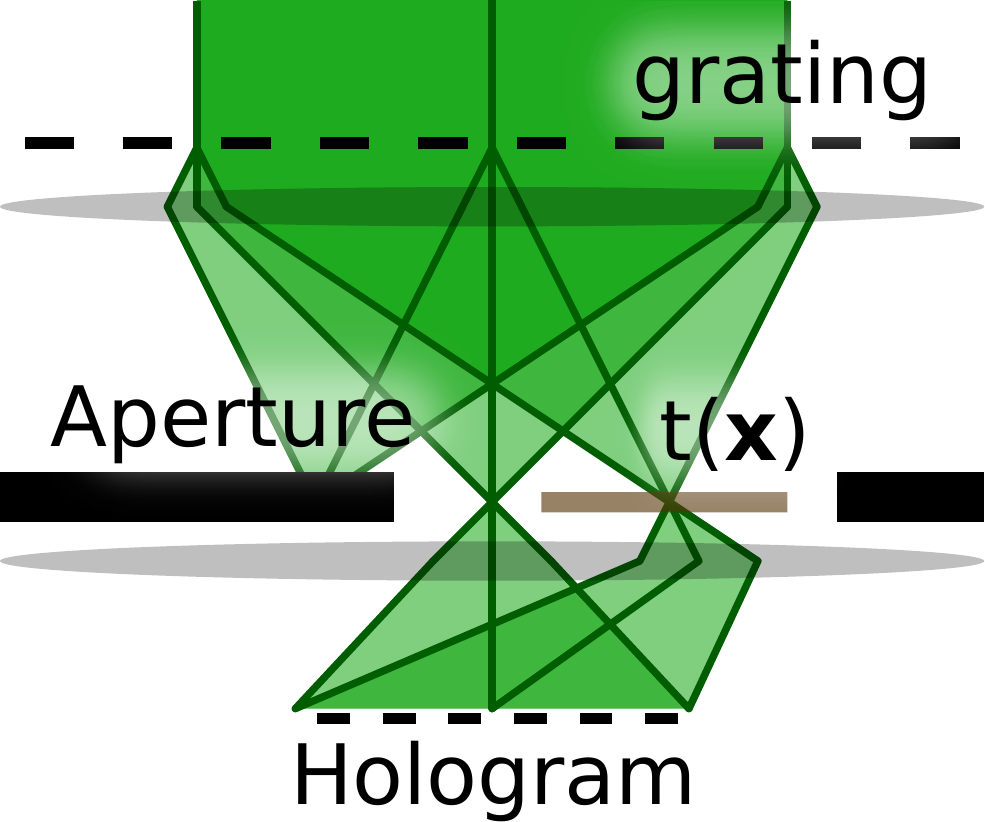}} \\
\caption{(a) Schematic of off-axis electron holography with a biprism. One plane wave is passed through the specimen (brown), and an electrostatic biprism (black dot) interferes this wave with a second plane wave passed through vacuum. (b) Schematic of STEMH. A diffraction grating in the condenser system produces multiple beams at the specimen (brown). An aperture (black) admits one beam that interacted with the specimen and one passed through vacuum. The projector system combines these beams into a hologram.\label{fig:schematic}} 
\end{figure}
  
Off-axis electron holography offers a more interpretable measurement of phase with respect to a vacuum reference wave. This allows, for example, imaging of magnetic bits in recording media \cite{osakabe_observation_1983} and insight into the charge distribution and asymmetry of nanoparticles \cite{kim_dipole-like_2018}. However, as interference fringes are in real space, resolution is limited by the spacing of fringes \cite{lichte_electron_1991}, coherence, and biprism stability. In this manuscript, we demonstrate a method to measure specimen-induced electron phase shifts measured from the interference of multiple STEM probe beams produced with an electron diffraction grating. Between 1989 and 1994, several groups proposed or demonstrated early forms of STEM holography (STEMH) using an electron biprism to produce two beams \cite{leuthner_STEM-holography_1989,cowley_high_1990,takahashi_observation_1994}; because the speed of pixelated detectors was insufficient at the time, these first demonstrations employed a grating mask to map fringe shifts into a single intensity signal per probe position.  % Other early work employed a STEM instrument without scanning and interference fringes in defocused real space \cite{mankos_absolute_1994,mankos_quantitative_1996}.  
Our implementation of the technique is different than these early approaches, but as the basic principle matches these works, we retain the same name.

An electron diffraction grating has several advantages over a biprism for STEMH. The coherence width necessary for optimal fringe visibility is much lower for an amplitude-dividing beamsplitter than a wavefront-dividing beamsplitter \cite{marton_electron_1952,marton_electron_1953,matteucci_amplitude_1981}. Furthermore, a biprism produces two opposing half-circular probes in reciprocal space, whereas a grating can produce probes with identical phase and amplitude distributions. The ability to tune the phase structure of each diffracted beam allows for versatile extensions of STEMH, including the possiblity to map out-of-plane magnetic fields \cite{grillo_observation_2017} (see Appendix section \ref{sect:structure}). Cowley proposed several methods to reconstruct phase in STEMH with exactly two probe beams \cite{cowley_high_1990,cowley_ultra-high_2003}. We previously demonstrated a three-beam STEM electron interferometer and proposed a method for reconstructing phase when the probe size is much smaller than specimen phase variations \cite{yasin_path-separated_2018}. In this manuscript, we start with a general approach to reconstruct phase from the two or more beams with tunable phase structure produced by a diffraction grating. We then treat the two-beam case in detail and demonstrate the approach in experiment.

%   \footnote{Indeed, identical probes are necessary to realize one of the reconstruction methods Cowley proposed. The optimal instrument designed to realize his proposal might be a STEM instrument with a diffraction grating in the condenser aperture and a second aperture at a crossover above the specimen to select only two beams.}
\section{Model and Reconstruction}

\subsection{General case}

First, we shall introduce our notation. We use a pre-specimen probe wavefunction
\begin{equation}
  \psi_i(\mathbf{x}) = a(\mathbf{x}-\mathbf{x}_p)
\end{equation}
where $\mathbf{x}_p$ is the offset-position of our probe. For thin specimens, we can describe the interaction of the probe with the specimen as a multiplication by a specimen transmission function $t(\mathbf{x})$, resulting in a post-specimen wavefunction
\begin{equation}
  \psi_f(\mathbf{x}) = a(\mathbf{x}-\mathbf{x}_p)t(\mathbf{x})
\end{equation}
and an interference pattern at the detector at probe position $\mathbf{x}_p$
\begin{equation} \label{eq:grating_image}
  I_p(\mathbf{k}) = \left|\psi_f(\mathbf{k})\right|^2_p = \left[A_p^{*}(\mathbf{k})\otimes T^{*}(\mathbf{k})\right]\left[A_p(\mathbf{k})\otimes T(\mathbf{k})\right]
\end{equation}
where $A_p(\mathbf{k})$ is the Fourier transform of $a(\mathbf{x}-\mathbf{x}_p)$ and $T(\mathbf{k})$ is the Fourier transform of $t(\mathbf{x})$.
Using a diffraction grating to produce multiple sharply-peaked, evenly spaced probes, our probe wavefunction is
\begin{equation}
  a(\mathbf{x}- \mathbf{x}_p) = \sum_m c_m a_m(\mathbf{x}-m \mathbf{x}_0 - \mathbf{x}_p),
\end{equation}
where $a_m(\mathbf{x})$ is sharply peaked at $\mathbf{x} = \mathbf{0}$ and thus
\begin{equation}
  A_p(\mathbf{k}) = \sum_m c_m e^{-i\mathbf{k}\cdot\left(m \mathbf{x}_0 +\mathbf{x}_p\right)} A_m(\mathbf{k}).
\end{equation}

If we plug this into \eqref{eq:grating_image}, and move the plane wave terms through the convolution \footnote{
\begin{equation*}
  \left(f(x) e^{i k x}\right) \otimes g(x) = \left[f(x) \otimes \left(g(x) e^{-i k x}\right) \right] e^{i k x}. 
\end{equation*}},
% \begin{equation} \label{eq:full_grating_image} 
%   G_p(\mathbf{k}) =  \sum_{m,n} c_m^* c_n\left[\left(A_m^{*}(\mathbf{k})e^{i\mathbf{k}\cdot\left(m \mathbf{x}_0 +\mathbf{x}_p\right)} \right)\otimes T^{*}(\mathbf{k})\right]\left[\left(A_n(\mathbf{k})e^{-i\mathbf{k}\cdot\left(n \mathbf{x}_0 +\mathbf{x}_p\right)}\right)\otimes T(\mathbf{k})\right]
% \end{equation}
we see that \eqref{eq:grating_image} can be rewritten as
% \begin{widetext}
\begin{align} 
  I_p(\mathbf{k}) =  \sum_{m,n} c_m^* c_n 
  \left[A_m^{*}(\mathbf{k}) \otimes 
  \left( T^{*}(\mathbf{k}) e^{-i\mathbf{k}\cdot\left(m \mathbf{x}_0 +\mathbf{x}_p\right)}\right)\right]\cdot \nonumber \\
  \left[A_n(\mathbf{k})\otimes 
  \left( T(\mathbf{k}) e^{i\mathbf{k}\cdot\left(n \mathbf{x}_0 +\mathbf{x}_p\right)}\right)\right] 
  e^{-i(n-m)\mathbf{k}\cdot \mathbf{x}_0 }. \label{eq:master}
\end{align}
% \end{widetext}
We can see that the specimen transmission function $t(\mathbf{x})$ is encoded in the set of plane waves $e^{-i(n-m)\mathbf{k}\cdot \mathbf{x}_0 }$.

If we take the inverse Fourier transform, we see that the plane waves in \eqref{eq:master} correspond to spatially separated spots.
%\begin{align} \label{eq:twobeam_master}
%  G_p(\mathbf{k}) =&  c_0^* c_1 
%  A_0^{*}(\mathbf{k})
%  \left[A_0(\mathbf{k})\otimes 
%  \left( T(\mathbf{k}) e^{i\mathbf{k}\cdot\left(\mathbf{x}_0 +\mathbf{x}_p\right)}\right)\right] 
%  e^{-i\mathbf{k}\cdot\mathbf{x}_0 } \nonumber \\
%  &+ c.c. + G_0(\mathbf{k},\mathbf{x}_p)
%\end{align}
%where $G_0$ corresponds to the $n=m$ terms that contain information only about the amplitude of the specimen transmission function. Let us take the Fourier transform of $G_p$.

\begin{figure}%[t]
  \includegraphics[width=\columnwidth]{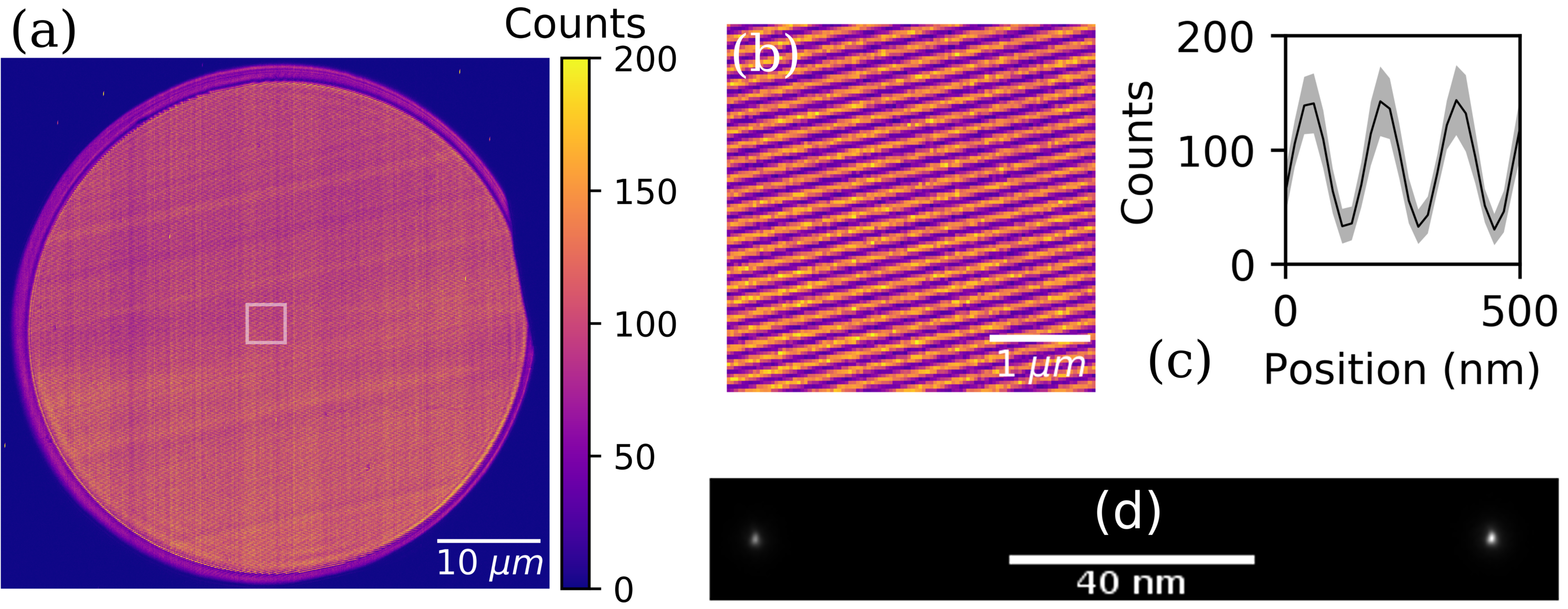}
%   \subfloat[]{\label{subfig:overview:grating}
%   \includegraphics[width=1.8in]{full_grating_nocblabel.png} }
%   \begin{minipage}{1.5in}
%     \vspace{-0.6in}
%     \begin{center}
%       \subfloat[]{\label{subfig:overview:grating_inset}
%       \includegraphics[width=0.8in]{grating_inset_nocb.png} } \\
%       \hspace{-0.1in}
%       \subfloat[]{\label{subfig:overview:grating_bars
%       \includegraphics[width=1.2in]{grating_bars.png} }
%     \end{center}
%   \end{minipage} \\
%   \begin{minipage}{0.7\columnwidth}
%     \subfloat[]{\label{subfig:overview:beams}
%     \includegraphics[height=0.6in]{beams_in_SA_uncalibrated_cropped.png}}
%   \end{minipage}
%   \begin{minipage}{0.2\columnwidth}
%     \vspace{0.06in}
%     \subfloat[]{\label{subfig:overview:fft}
%     \includegraphics[height=0.68in]{ref_beam_1_scaled.png} }
%   \end{minipage} 
  \caption{(a) Measured interference fringes formed by two beams in vacuum. (b) Zoom-in of the region in (a) highlighted by a white rectangle (same colorbar). (c) Line profile (black) with 95\% confidence interval (grey) of interference fringes in the center of (a). (c) Micrograph of beams used for experiment. The beam separation is $|\mathbf{x}_0| = \unitmo{120}{nm}$. \label{fig:overview}}
  %(d) Diffracted beams passed through selected area aperture in vacuum. (e) Amplitude (brightness) and phase (color) of the first order in the Fourier transform of (a). \label{fig:overview}}
\end{figure}

\begin{equation}
	\littleI (\mathbf{x}_p,\mathbf{x}) = \frac{1}{2\pi}\int \mathrm{d}\mathbf{k}\ e^{i\mathbf{k}\cdot\mathbf{x}} I_p(\mathbf{k}) = \sum_{\ell} \littleI_{\ell}(\mathbf{x}_p,\mathbf{x}),
\end{equation}
where $\ell = n-m$ and each $\littleI_{\ell}$ term contains a sum over $m$. For example, $\littleI_0$ corresponds to the $n=m$ terms that contain information only about the amplitude of the specimen transmission function, and $\littleI_{-1}$ contains the same information as $\littleI_1$. Each $\littleI_{\ell}(\mathbf{x}_p,\mathbf{x})$ is sharply peaked at $\mathbf{x}=\ell\mathbf{x}_0$.
We can get better insight into the information encoded here if we restrict our attention to a limited set of plane waves.
% How do we retrieve it? We will consider a set of simiplifying assumptions to illustrate what \eqref{eq:master} means.

\subsection{Two beams, vacuum reference}

\begin{figure}
  \includegraphics[width=\columnwidth]{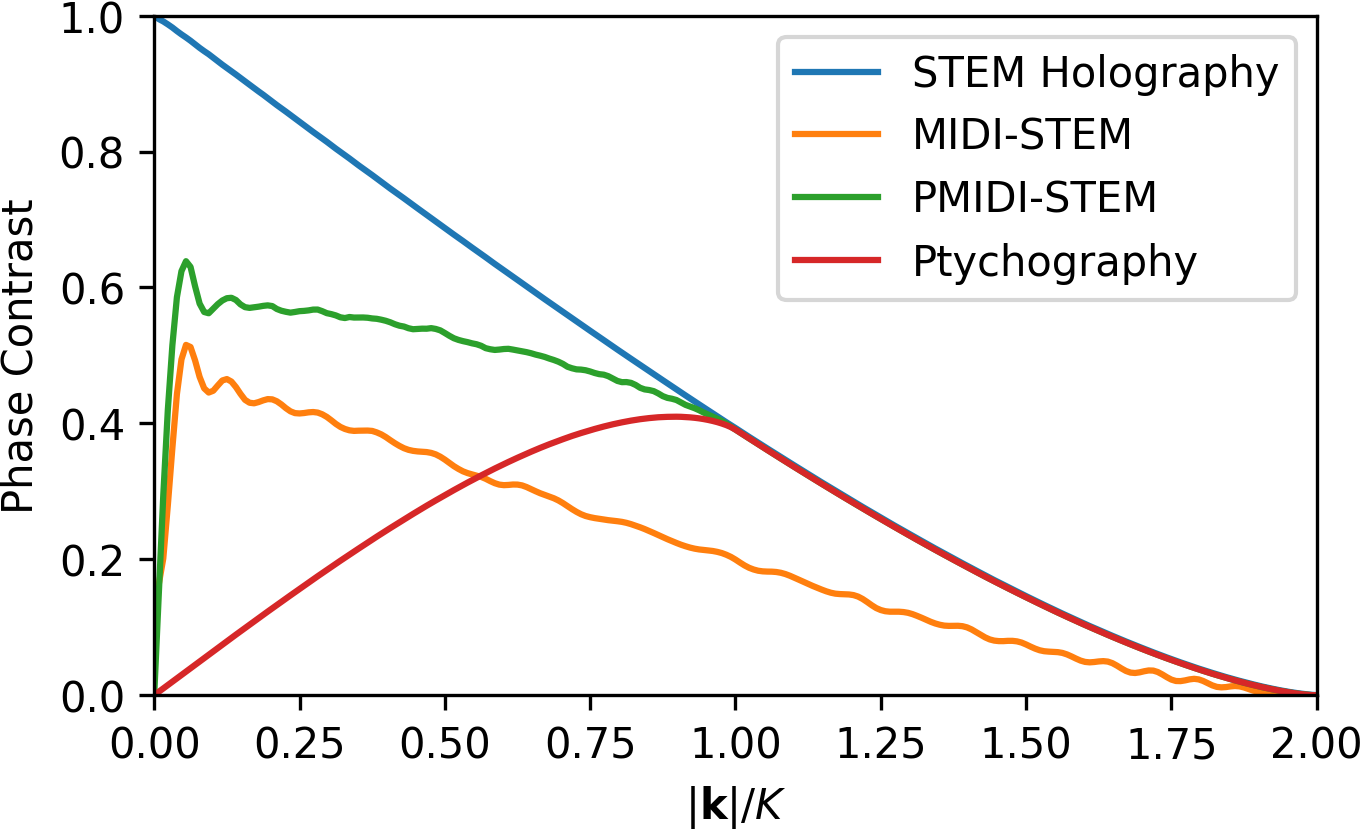}
  \caption{Comparison of calculated phase contrast transfer functions for several phase-contrast STEM techniques. Unlike MIDI-STEM \cite{ophus_efficient_2016}, PMIDI-STEM \cite{yang_enhanced_2016}, and pytchography \cite{yang_simultaneous_2016}, STEMH produces efficient contrast as the spatial frequency approaches zero (see \eqref{eq:t_M}).\label{fig:pctf}}
\end{figure}

In off-axis electron holography, only two plane waves interfere. The resulting fringe pattern is therefore straightforward to analyze. We can perform a similarly straightforward STEMH experiment by introducing an aperture to block extra beams. We will consider the case where all beams but the $m=0$ and $m=+1$ beams are blocked, and $m=0$ passes through vacuum, i.e. $t(\mathbf{x}) = 1$ near $\mathbf{x} = 0$ over a region larger than the maximum range of $\mathbf{x}_p$. We shall further assume that the aperture function $A_m = A_0$ is the same for both diffraction orders, and that it takes the form of a uniform disk,
\begin{equation}
  A_0(\mathbf{k}) = \begin{cases}
	  \frac{1}{\sqrt{\pi K^2}} & |\mathbf{k}| \leq K \\
    0 & |\mathbf{k}| > K
  \end{cases} \label{eq:aperture}
\end{equation}
where $K$ is the edge of the aperture. 

Now, with just two beams, only $\littleI_{-1}$, $\littleI_{0}$, and $\littleI_{1}$ are nonzero, and each one contains only a single term from the sum on $m$.
For clarity, we will take $\mathbf{x} \to \mathbf{x}+\mathbf{x}_0$, and therefore shift $\littleI_{1}$ into the center. This eliminates a delta function from $\littleI_1$. We can ignore $\littleI_0$ and $\littleI_{-1}$; we will choose a window for our integration that only includes the term we just moved to the center. In this work, we used a window with width $0.08|\mathbf{x}_0|$. Because one beam is in vacuum, $t(\mathbf{x}) = 1$ there. We therefore see that
% inverse discrete Fourier transform that only includes the term we just moved to the center.

\begin{equation}
  \littleI_{1}(\mathbf{x}_p,\mathbf{x})= c_0^* c_1 a_0(\mathbf{x}) \otimes \left[a_0(\mathbf{x}) t(\mathbf{x}+\mathbf{x}_0 + \mathbf{x}_p)\right]. \label{eq:doublereal}
\end{equation}

Although an iterative method to reconstruct $t(\mathbf{x})$ from \eqref{eq:doublereal} is possible, we focus here on the simpler linear reconstruction. This linear reconstruction could later serve as an initial guess for an iterative reconstruction. We want an interpretable function of just $\mathbf{x}_p$, but we have two position variables. The simplest way to trace one out is to integrate over $\mathbf{x}$ with $a_0(\mathbf{x})$ as a kernel.

% This integral is very simple, as $a_0(\mathbf{x}) \otimes a_0(\mathbf{x}) = a_0(\mathbf{x})$

\begin{align} 
	t_M(\mathbf{x}_p+\mathbf{x}_0) &= -\frac{\sqrt{\pi K^2}}{c_0^* c_1} \int \mathrm{d}\mathbf{x} \  
                        a_0(\mathbf{x}) \littleI_1(\mathbf{x}_p,\mathbf{x})  \nonumber \\
%   &= \int \mathrm{d}\mathbf{x} \int \mathrm{d}\mathbf{x}' c_0^* c_1 a_0(\mathbf{x}) a_0(\mathbf{x}-\mathbf{x}') \left[a_0(\mathbf{x}') t(\mathbf{x}'+\mathbf{x}_0 + \mathbf{x}_p)\right] \\
%   &= \int \mathrm{d}\mathbf{x}' c_0^* c_1 |a_0(\mathbf{x}')|^2 t(\mathbf{x}'+\mathbf{x}_0 + \mathbf{x}_p) \\
  &=  h(\mathbf{x}_p) \otimes 
                  t(\mathbf{x}_p+\mathbf{x}_0), \label{eq:t_M}
\end{align}

We see that the object we have defined, which we call $t_M$ to mean the measured transfer function, is exactly the specimen transmission function convolved with a point spread function $h(\mathbf{x}_p) = |a_0(\mathbf{x}_p)|^2$ for our choice of aperture function \eqref{eq:aperture} 
\footnote{
  A round aperture function with no phase structure has the property that $a_0(\mathbf{x}) \otimes a_0(\mathbf{x}) = \frac{1}{\sqrt{\pi K^2}} a_0(\mathbf{x})$. See Appendix section I.
}. This result matches that of Cowley \cite{cowley_high_1990}, even though the reconstruction method differs.
%(see Supplementary Section ((((())))) for a more detailed comparison). 
A less trivial aperture function only changes $h(\mathbf{x}_p)$ in \eqref{eq:t_M}, as we show in Appendix sections \ref{sect:aberrations} and \ref{sect:structure}. We can get a little more insight into the effect of this point spread function on our image by looking at its reciprocal space equivalent, the contrast transfer function (CTF) \cite{kirkland_advanced_2010}. 

The contrast transfer function measures the efficiency with which an imaging method reconstructs the spatial frequencies which make up an image. The nearly-unity efficiency of STEMH as the spatial frequency approaches zero, shown in Figure \ref{fig:pctf}, is a unique feature of STEMH. Unlike existing phase-contrast STEM techniques, where the value of the reconstructed phase of any one pixel is meaningful only with respect to its neighbors in some finite-sized region, in STEMH, the phase recorded in one pixel offers a meaningful comparison to an electron passed through vacuum. This means that STEMH could be used to quantitatively measure thickness or electric and magnetic fields. 
%, as long as phase differences exceeding $2\pi$ are appropriately unwrapped. 

We will show next that STEMH and the measurement method we have outlined can be employed to image nanoparticles, carbon substrates and contamination, and electric fields.

\section{Experiment}

\begin{figure}[h]
  \subfloat[]{\label{subfig:comparison:amp12}
  \includegraphics[height=0.95in]{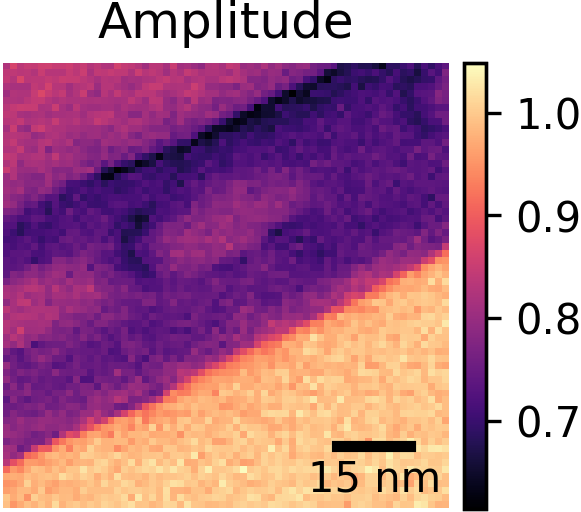} }
  \subfloat[]{\label{subfig:comparison:phase12}
  \includegraphics[height=0.95in]{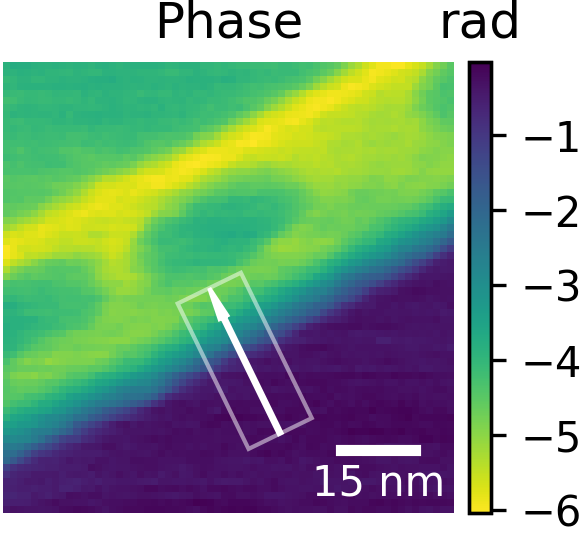} }
  \subfloat[]{\label{subfig:comparison:haadf12}
  \includegraphics[height=0.95in]{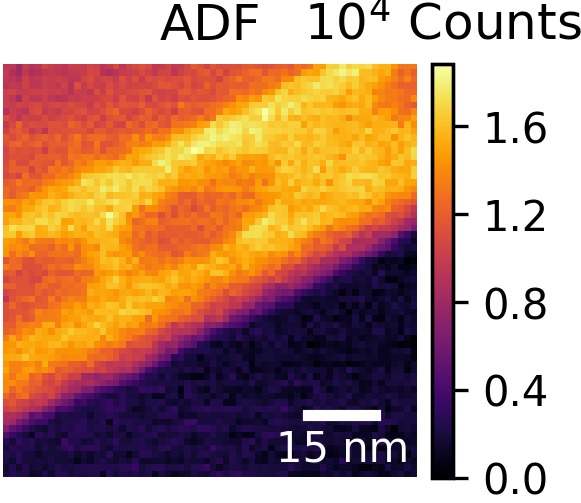} } \\
  \subfloat[]{\label{subfig:comparison:amp17}
  \includegraphics[height=0.85in]{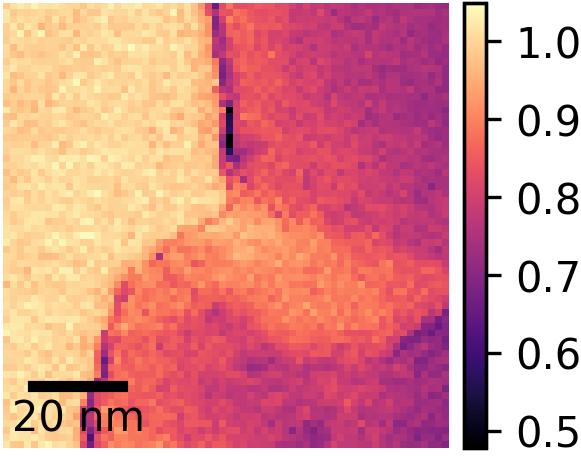} }
  \subfloat[]{\label{subfig:comparison:phase17}
  \includegraphics[height=0.85in]{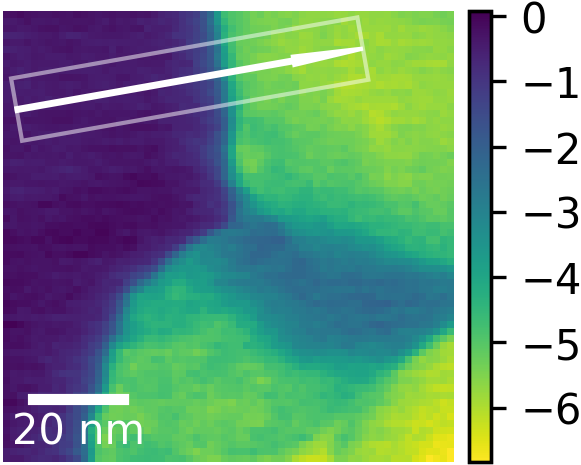} }
  \subfloat[]{\label{subfig:comparison:haadf17}
  \includegraphics[height=0.85in]{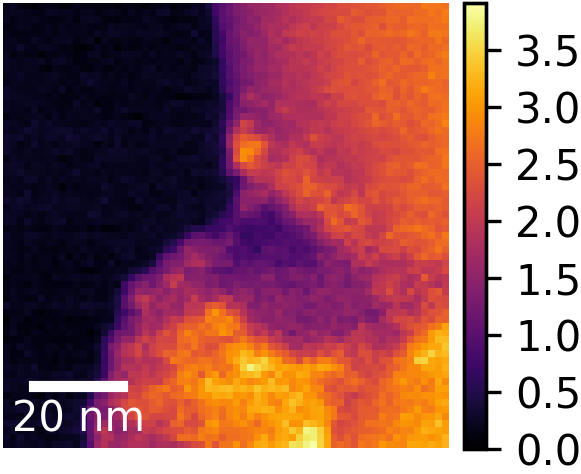} } \\
  \subfloat[]{\label{subfig:comparison:lacey_lineplot}
  \includegraphics[height=1.1in]{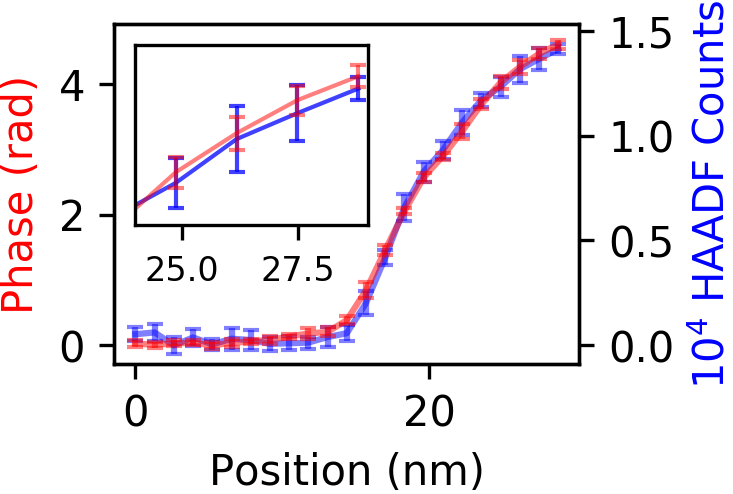} }
  \subfloat[]{\label{subfig:comparison:CdTe_lineplot}
  \includegraphics[height=1.1in]{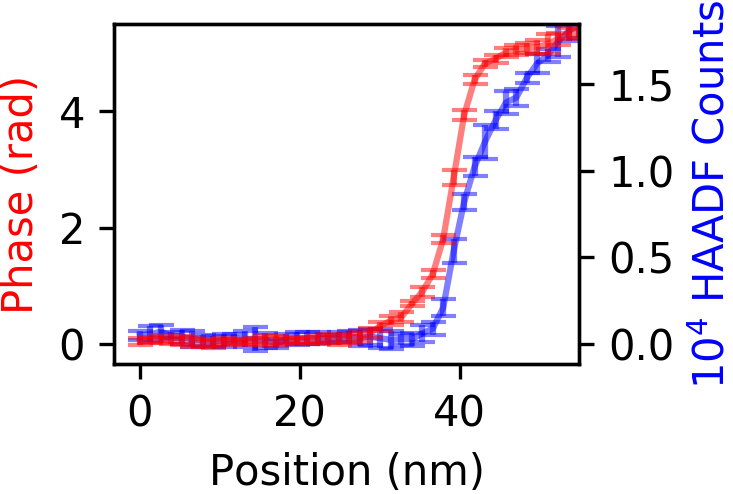} }\\
  \caption{Comparison of micrographs recorded by STEMH and ADF-STEM on lacey carbon and semidconducting nanoparticles. (a,d) Amplitude measured by STEMH. (b,e) Phase measured by STEMH; line profiles in (g,h) are taken along the white arrow and averaged over the width of the box. (c,f) Simultaneously acquired ADF. (g) Comparison of phase (red) with ADF (blue) on lacey carbon edge. Inset: zoom-in to show noise levels. (h) Same as (g) on the edge of a semiconducting nanoparticle. \label{fig:comparison}}%Inset: zoomed-in plot of the vacuum region just outside the particle. The phase outside the particle is fit well by an exponential decay (green). \label{fig:comparison}}
  % starting at the edge of the particle, as determined by the maximum gradient of the ADF. 
\end{figure}

% TO-DO: CdTe information

We tested our implementation of STEMH on three types of specimens: a lacey carbon substrate, gold nanoparticles on lacey and ultrathin carbon, and semiconducting CdTe/CdS nanoparticles synthesized according to procedures described previously \cite{tang_spontaneous_2002,lee_bioconjugated_2006}.

In order to produce multiple diffracted beams, we placed a \unittw{50}{\mu}{m}-diameter diffraction grating with a \unitto{150}{nm} pitch (see Fig. \ref{fig:overview}a) in the third condenser aperture strip of the TEAM I microscope at Lawrence Berkeley National Laboratory. The second condenser aperture was used to block transmission outside the grating. As the grating was partially blazed, the amplitudes \footnote{Normalized to the total counts transmitted through the selected area aperture in vacuum.} of the zeroth- and first-order beams were $c_0 = 0.79$ and $c_1 = 0.61$, respectively (see Fig. \ref{fig:overview}d). We measure an interference fringe visibility $V = 70.7\%$. As the maximum possible for our measured beam amplitudes $c_0$ and $c_1$ is $V = 96.7\%$, our measured value is likely lower due to inelastic scattering in the grating, aberrations in the projector lens system, and an imperfect detector modulation transfer function (MTF).

For a first test of our reconstruction method, we chose a \unitto{4}{mrad} convergence semi-angle so that we could easily block all but two beams with the selected area aperture. This is also possible with a higher convergence angle, but may require a custom-made smaller aperture to cleanly block other beams. The phase measured by STEMH is insensitive to diffraction at this convergence angle, as diffracted disks do not overlap with the center disk.% and are in fact outside of the detector in this experiment. 
We performed experiments on the TEAM I microscope with an incident electron energy of \unitto{300}{keV}, and recorded data with the Gatan K2 IS direct electron detector at 400 fps with a camera length $L = \unitmo{1.45}{m}$. % Because the diffraction defocus is non-zero with factory alignment of the projector system on an FEI Titan in STEM, we tested two configurations of the projector system: (i) no adjustment of the diffraction defocus and a camera length $L=\unitmo{1.8}{m}$, and (ii) minimized defocus and a camera length of $L = \unitmo{1.45}{m}$. Configuration (i) comes with a lower efficiency, as the two beams are not fully overlapped. The descan is less well-aligned in configuration (ii), and largerscan-induced shifts of the diffraction disk are posssible.

We found that the phase measured by STEMH and the ADF-STEM signal agreed very well on lacey carbon, as shown in Fig. \ref{subfig:comparison:lacey_lineplot}. Since lacey carbon has no diffraction contrast and is conductive, both techniques produce mass-thickness contrast.

However, on and near a semiconducting nanoparticle, charging of the particle strongly affects the phase and does not affect the ADF, as shown in Fig. \ref{subfig:comparison:CdTe_lineplot}. %The profile of the phase in vacuum outside the edge of the particle, as determined by the maximum gradient of the ADF, is fit well by an exponential decay. 
The organic stabilizers used in the synthesis of the nanoparticles may persist on the surface and contribute to charging \cite{kim_dipole-like_2018}. There are clusters on the surface of the particles which may be electrically insulated. The clusters can be seen most clearly in the ADF (Fig. \ref{subfig:comparison:haadf17}). As the clusters do not stand out in the phase image (Fig. \ref{subfig:comparison:phase17}), it is likely that the average atomic number of the clusters is close to that of the particles but crystalline order produces diffraction contrast. 

\begin{figure}%[h]
  \begin{tabular}{l l}
    \subfloat[]{\label{subfig:Capt3:amp}
    \includegraphics[height=0.65in]{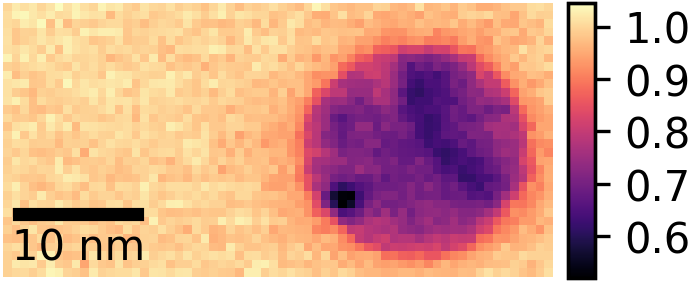} } &
    \subfloat[]{\label{subfig:Capt3:phase}
    \includegraphics[height=0.65in]{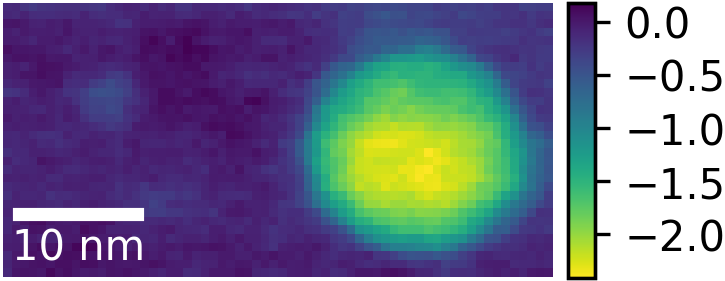} } \\
    \begin{minipage}{1.5in}
    \vspace{-0.7in}
    \subfloat[]{\label{subfig:Capt3:ph-amp}
    \includegraphics[height=0.85in]{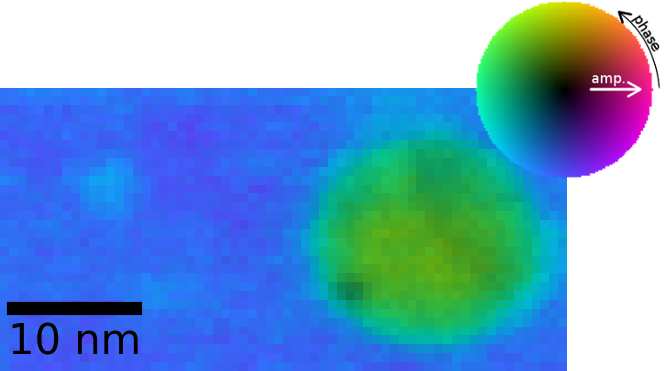} } \end{minipage}
    &
    \subfloat[]{\label{subfig:Capt3:haadf}
    \includegraphics[height=0.7in]{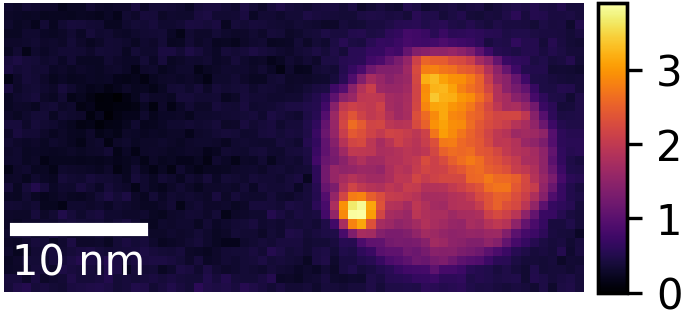} } \\
  \end{tabular}
  \caption{Au nanoparticle with reference beam on uniform ultrathin carbon substrate. (a) Amplitude from STEMH. (b) Phase from STEMH. (c) Phase and amplitude (see colorwheel) shown together offers more information than either alone. (d) Simultaneously acquired ADF signal. \label{fig:Capt3}}
\end{figure}

Imaging with STEMH also works with the reference beam on a uniform substrate if no vacuum region is accessible. In Fig. \ref{fig:Capt3}, we passed the reference beam through ultrathin carbon and scanned the imaging beam over a gold nanoparticle on ultrathin carbon. Increased noise is likely with high doses on a reference area of a uniform substrate, as STEMH is highly sensitive to deposited contamination. The spot that is barely distinguishable from noise in the upper left of the ADF image, Fig. \ref{subfig:Capt3:haadf}, is quite clear in the phase image, Fig. \ref{subfig:Capt3:phase}. So, if the reference beam deposits contamination, the imaged phase will include contributions from the uncharacterized phase of the reference area, which is effectively noise. 

Although the phase measured by STEMH may typically be more useful, the amplitude also offers valuable information. The amplitude image is similar to a bright field image, with linear rather than quadratic sensitivity to amplitide changes in the bright field disk. With our low convergence angle, diffraction produces only amplitude contrast. With a higher convergence angle, interference of the center disk and diffracted disks produces both amplitude and phase contrast. Strong linear phases from particle edges cause a shift in diffraction and therefore a reduced overlap of the two disks, so the amplitude image has good edge contrast. A combined phase-amplitude image is sometimes more interpretable than either alone, as seen in Fig. \ref{subfig:Capt3:ph-amp}.

\begin{figure}%[h]
  \begin{tabular}{l l}
    \subfloat[]{\label{subfig:sim:amp}
    \includegraphics[width=1.5in]{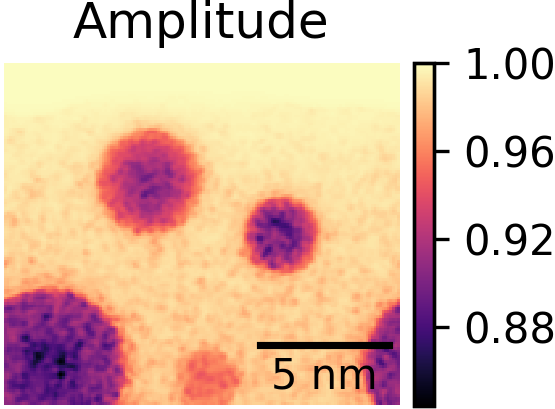} } &
    \subfloat[]{\label{subfig:sim:phase}
    \includegraphics[width=1.5in]{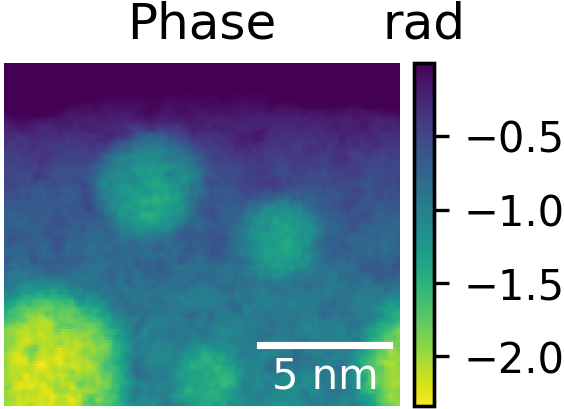} } \\
    \begin{minipage}{1.35in}
      \vspace{0.25in}
    \subfloat[]{\label{subfig:sim:haadf}
    \includegraphics[width=1.35in]{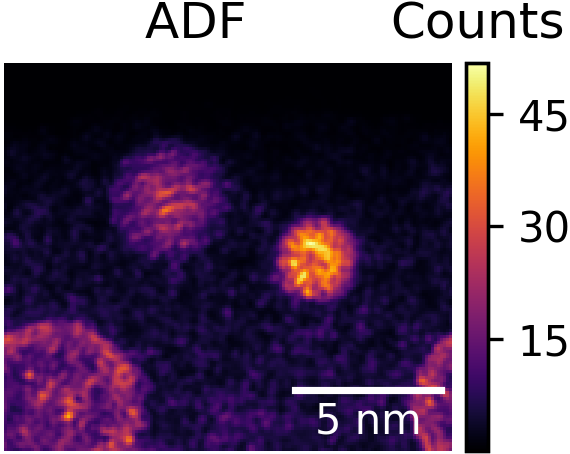} } 
    \end{minipage} &
    \begin{minipage}{1.30in}
      \vspace{0.0in}
    \subfloat[]{\label{subfig:sim:projpot}
    \includegraphics[width=1.30in]{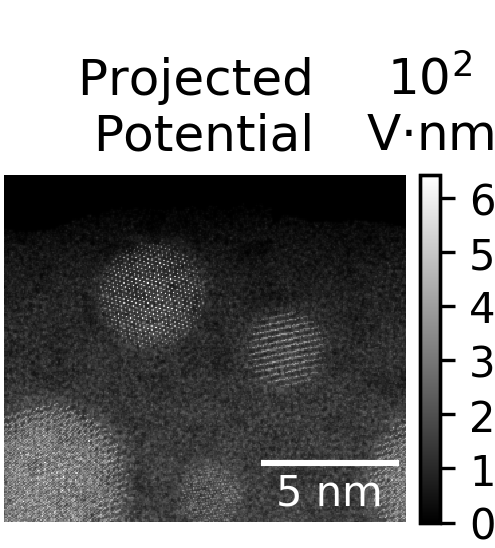} } 
    \end{minipage} \\
  \end{tabular}
  \caption{Simulated STEMH dataset with Au nanoparticles on a carbon wedge and reference beam in vacuum. (a) Amplitude. (b) Phase.  (c) ADF from the same dataset. (d) Projected potential used to generate the dataset. \label{fig:2beamsim}}
\end{figure}

Simulations of a STEMH experiment with beam separation $x_0 = \unit[15]{\textrm{nm}}$ and a convergence semi-angle of $\unit[4]{\textrm{mrad}}$, on gold particles embedded in an amorphous carbon wedge support our experimental observations \cite{ricolleau_random_2013}. The prism algorithm \cite{ophus_fast_2017} implemmented in the Prismatic code \cite{pryor_streaming_2017}, was used to produce each probe simulation, which were combined coherently in the far field to form STEMH diffraction patterns. The phase \ref{subfig:sim:phase} more clearly matches the projected potential \ref{subfig:sim:projpot} than the ADF signal \ref{subfig:sim:haadf}, as contrast is much stronger on the carbon wedge. See Appendix section \ref{sect:sim} for more detailed comparison of phase and projected potential. As we used an ADF detector inner semi-angle of $\unit[8]{\textrm{mrad}}$, we see diffraction contrast in both the ADF and amplitude signals. 

We see from experimental tests and comparison with simulation that STEMH offers efficient contrast on low- and high-atomic number materials as well as on electric fields. In particular, like ADF-STEM, but unlike other phase contrast STEM techniques, the contrast transfer function does not go to zero at zero spatial frequency (Fig. \ref{fig:pctf}), so efficient, quantitative thickness or electric and magnetic field measurement may be possible.

\section{Conclusion}

We have demonstrated a straightforward method to measure the transfer function of a specimen with STEMH. Unlike existing phase contrast STEM techniques, STEMH measures phase with respect to a vacuum reference. A single pixel in STEMH therefore has an absolute meaning, just as in HAADF-STEM and off-axis holography. STEMH also produces a lower-noise image than ADF (see Fig. \ref{subfig:comparison:lacey_lineplot}), with much better contrast on weak phase objects and sensitivity to electric and magnetic fields.

STEMH also has the potential to achieve higher resolution than off-axis electron holography. Unlike electron holography, the fringe spacing does not affect resolution in STEMH, since the fringes are in diffraction space. The real-space resolution is limited only by aberrations, coherence, and convergence angle of the probe. In Appendix section \ref{sect:aberrations}, we derive the measured transfer function $t_M(\mathbf{x}_p)$ in the presence of aberrations. We show in related work that atomic resolution phase measurement is possible with a higher convergence angle and aberration correction \cite{yasin_probing_2018}. Aberrations can be corrected holographically with the grating \cite{linck_aberration_2017}. %We will explore the possibility to correct aberrations with the choice of an appropriate kernel in future work.

It is worth exploring in detail in the future two more involved reconstruction methods: first, retrieving spatial information by iteratively updating the transfer function based on $\littleI_1(\mathbf{x}_p,\mathbf{x})$ rather than simply integrating out $\mathbf{x}$; and second, extending the field of view by treating scans where all beams begin in vacuum and all end on a specimen, using information from the first beam's interaction at position $\mathbf{x}_p$ to correct for the next beam's interaction at $\mathbf{x}_p$ and produce a flat phase reference for the first beam at position $\mathbf{x}_p + \mathbf{x}_0$. Both methods, while more computationally intensive, could significantly improve the utility of STEMH in understanding the fine structural details of cells, organic semiconductor interfaces, and nanostructures.

This simple modification to the electron microscope column--replacement of one condenser aperture with a diffraction grating--and straightforward reconstruction has the potential for versatile and efficient imaging. STEMH has sensitivity to electric and magnetic fields like off-axis holography and can image with a resolution comparable to ADF-STEM. 

\section{acknowledgments}

T.R.H. performed work at the Molecular Foundry with support from the the U.S. Department of Energy, Office of Science, Office of Workforce Development for Teachers and Scientists, Office of Science Graduate Student Research (SCGSR) program. The SCGSR program is administered by the Oak Ridge Institute for Science and Education for the DOE under contract number DE‐SC0014664. 
Work at the Molecular Foundry was also supported by the Office of Science, Office of Basic Energy Sciences, of the U.S. Department of Energy under Contract No. DE-AC02-05CH11231. 
F.S.Y. acknowledges support from the National Science Foundation Graduate Research Fellowship Program under Grant No. 1309047.
T.R.H., F.S.Y., J.J.C., J.S.P. and B.J.M. acknowledge support from the U.S. Department of Energy, Office of Science, Basic Energy Sciences, under Award DE-SC0010466.
R.M.S.dR. and J.C. acknowledge support from the U.S. Department of Energy Early Career Research Program.
W.F. and N.A.K. acknowledge funding from the National Science Foundation under Grant No. 1463474.
We thank the NVIDIA Corporation for donation of GPU resources.

\appendix 
\begin{widetext}
\section{Convolution of aperture functions \label{sect:a_convolve}}

It is straightforward to show that the convolution of two aperture functions in real space, in the absence of a phase on $A_0(\mathbf{k})$, is equal the aperture function with a normalization constant.

\begin{align}
a_0(\mathbf{x}) \otimes a_0(\mathbf{x}) &= \int \mathrm{d}\mathbf{x}' a_0(\mathbf{x}') a_0(\mathbf{x}-\mathbf{x}') \\
				     &= \frac{1}{(2\pi)^2}\int \mathrm{d}\mathbf{x}' \int \mathrm{d}\mathbf{k} e^{i\mathbf{k}\cdot \mathbf{x}'} A_0(\mathbf{k}) \int \mathrm{d}\mathbf{k}' e^{i\mathbf{k}'\cdot \left(\mathbf{x}-\mathbf{x}'\right)} A_0(\mathbf{k}') \\
				     & = \frac{1}{2\pi}\int \mathrm{d}\mathbf{k} A_0(\mathbf{k}) \int \mathrm{d}\mathbf{k}' e^{i\mathbf{k}'\cdot \mathbf{x}} A_0(\mathbf{k}') \delta(\mathbf{k}-\mathbf{k}') \\
				     & = \frac{1}{2\pi}\int \mathrm{d}\mathbf{k} |A_0(\mathbf{k})|^2 e^{i\mathbf{k}\cdot \mathbf{x}} \\
				     & = \frac{1}{\sqrt{\pi K^2}} \frac{1}{2\pi}\int \mathrm{d}\mathbf{k} A_0(\mathbf{k}) e^{i\mathbf{k}\cdot \mathbf{x}} \\
				     & = \frac{1}{\sqrt{\pi K^2}} a_0(\mathbf{x}). 
\end{align}

\section{Aberrations \label{sect:aberrations}}

It is worthwhile to consider the effect of aberrations on resolution in STEM holography. To do this, let us re-evalutate the measured transfer function $t_M(\mathbf{x}_p+\mathbf{x}_0)$ \eqref{eq:t_M} with aberrations included in $a_a(\mathbf{x})$, i.e. use instead
\begin{equation}
	a_a(\mathbf{x}) = \frac{1}{2\pi} \int \mathrm{d}\mathbf{k} A_0(\mathbf{k}) e^{i \chi(\mathbf{k})} e^{i\mathbf{k}\cdot \mathbf{x}}
\end{equation}
and use a kernel
\begin{equation}
	a^*_0(\mathbf{x}) = \frac{1}{2\pi} \int \mathrm{d}\mathbf{k} A_0(\mathbf{k}) e^{-i\mathbf{k}\cdot \mathbf{x}}.
\end{equation}

With these definitions, \eqref{eq:t_M} becomes
\begin{equation}
	t_M(\mathbf{x}_p+\mathbf{x}_0) = -\sqrt{\pi K^2} \int \mathrm{d}\mathbf{x} a^*_0(\mathbf{x}) \left(a^*_a(-\mathbf{x}) \otimes \left[a_a(\mathbf{x})t(\mathbf{x}+\mathbf{x}_0+\mathbf{x}_p)\right]\right).
\end{equation}
We have retained complex conjugation and a sign-flip on $a^*_a(-\mathbf{x})$ that we dropped for $a_0(\mathbf{x})$ by symmetry in \eqref{eq:t_M}.
If we now insert the definitions for $a^*_c(\mathbf{x})$ and $a_a(\mathbf{x})$ and write out the convolution, we see that
\begin{align}
	t_M(\mathbf{x}_p+\mathbf{x}_0) &=  -\frac{\sqrt{\pi K^2}}{(2\pi)^3} \int \mathrm{d}\mathbf{x} \mathrm{d}\mathbf{x}' \int \mathrm{d}\mathbf{k} \mathrm{d}\mathbf{k}' \mathrm{d}\mathbf{k}'' 
  A_0(\mathbf{k}) e^{-i\mathbf{k}\cdot \mathbf{x}} 
  A_0(\mathbf{k}') e^{-i\chi(\mathbf{k}')} e^{i\mathbf{k}'\cdot (\mathbf{x}-\mathbf{x}')} 
  A_0(\mathbf{k}'') e^{i\chi(\mathbf{k}'')} e^{i\mathbf{k}''\cdot \mathbf{x}'} 
  t(\mathbf{x}'+\mathbf{x}_0+\mathbf{x}_p) \\
	&= -\frac{\sqrt{\pi K^2}}{(2\pi)^2} \int \mathrm{d}\mathbf{x}' \int \mathrm{d}\mathbf{k} \mathrm{d}\mathbf{k}' \mathrm{d}\mathbf{k}'' \delta(\mathbf{k}'-\mathbf{k}) 
  A_0(\mathbf{k}) A_0(\mathbf{k}') e^{-i\chi(\mathbf{k}')} e^{-i\mathbf{k}'\cdot \mathbf{x}'} 
  A_0(\mathbf{k}'') e^{i\chi(\mathbf{k}'')} e^{i\mathbf{k}''\cdot \mathbf{x}'} 
  t(\mathbf{x}'+\mathbf{x}_0+\mathbf{x}_p) \\
	&= -\frac{\sqrt{\pi K^2}}{(2\pi)^2} \int \mathrm{d}\mathbf{x}' \int \mathrm{d}\mathbf{k} \mathrm{d}\mathbf{k}'' 
  A_0(\mathbf{k})^2 e^{-i\chi(\mathbf{k})} e^{-i\mathbf{k}\cdot \mathbf{x}'} 
  A_0(\mathbf{k}'') e^{i\chi(\mathbf{k}'')} e^{i\mathbf{k}''\cdot \mathbf{x}'} 
  t(\mathbf{x}'+\mathbf{x}_0+\mathbf{x}_p) \\
	&= -\frac{1}{(2\pi)^2} \int \mathrm{d}\mathbf{x}' \int \mathrm{d}\mathbf{k} \mathrm{d}\mathbf{k}'' 
  A_0(\mathbf{k}) e^{-i\chi(\mathbf{k})} e^{-i\mathbf{k}\cdot \mathbf{x}'} 
  A_0(\mathbf{k}'') e^{i\chi(\mathbf{k}'')} e^{i\mathbf{k}''\cdot \mathbf{x}'} 
  t(\mathbf{x}'+\mathbf{x}_0+\mathbf{x}_p) \\
	&= \int \mathrm{d}\mathbf{x}' 
	|a_a(-\mathbf{x'})|^2
  t(\mathbf{x}_0+\mathbf{x}_p-\mathbf{x}').
\end{align}
We now see that, with an aberrated probe $a_a(\mathbf{x})$, 
\begin{equation} \label{eq:aberrated_t_M}
  t_M(\mathbf{x}_p+\mathbf{x}_0) = |a_a(-\mathbf{x}_p)|^2 \otimes t(\mathbf{x}_0+\mathbf{x}_p).
\end{equation}
As is the case for incoherent imaging in STEM, aberrations just produce a point-spread function $h(\mathbf{x}) = |a_a(-\mathbf{x})|^2$, the probe shape with aberrations, that is larger than the aberration-free point-spread function $|a_0(\mathbf{x})|^2$. We have kept the sign in $|a_a(-\mathbf{x})|^2$ for generality in the case of asymmetric probe shapes. The cubic phase associated with coma, for example, produces a probe for which $h(\mathbf{x}) \neq h(-\mathbf{x})$.

\section{Structured probes \label{sect:structure}}

The measured transfer function $t_M(\mathbf{x}_p+\mathbf{x}_0)$ is also not strongly impacted by the inclusion of phase structure in the diffraction grating to produce a structured probe, such as a vortex beam \cite{verbeeck_production_2010,mcmorran_electron_2011,grillo_highly_2014,harvey_efficient_2014}. In this case, $A_m(\mathbf{k})$ includes an additional phase term,
\begin{equation} \label{eq:A_m}
  A_m(\mathbf{k}) = A_0(\mathbf{k}) e^{i \chi_m(\mathbf{k})},
\end{equation}
and typically $\chi_0 = 0$. If an aperture is used to pass the $m=0$ and $m=1$ beams, we record a fringe pattern that includes the term
\begin{equation}
  \littleI_1(\mathbf{x}_p,\mathbf{x}) = c_0^*c_1 a_0(\mathbf{x}) \otimes \left[ a_1(\mathbf{x}) t(\mathbf{x} + \mathbf{x}_0 +\mathbf{x}_p) \right].
\end{equation}
If we now employ a kernel with matched structure $a_1^*(\mathbf{x})$, we see that
\begin{align}
	t_M(\mathbf{x}_p+\mathbf{x}_0) &= -\sqrt{\pi K^2} \int \mathrm{d}\mathbf{x} a_1^*(\mathbf{x}) \left(a_0(\mathbf{x}) \otimes \left[ a_1(\mathbf{x}) t(\mathbf{x} + \mathbf{x}_0 +\mathbf{x}_p) \right] \right) \\
	&= -\frac{\sqrt{\pi K^2}}{(2\pi)^3} \int \mathrm{d}\mathbf{x} \mathrm{d}\mathbf{x}' \int \mathrm{d}\mathbf{k} \mathrm{d}\mathbf{k}' \mathrm{d}\mathbf{k}'' 
  A_0(\mathbf{k}) e^{-i \chi_1(\mathbf{k})} e^{-i\mathbf{k}\cdot \mathbf{x}} 
  A_0(\mathbf{k}') e^{i\mathbf{k}'\cdot (\mathbf{x}-\mathbf{x}')} 
  A_0(\mathbf{k}'') e^{i\chi_1(\mathbf{k}'')} e^{i\mathbf{k}''\cdot \mathbf{x}'}
  t(\mathbf{x}'+\mathbf{x}_0+\mathbf{x}_p) \\
  &= -\frac{\sqrt{\pi K^2}}{(2\pi)^2} \int \mathrm{d}\mathbf{x}' \int \mathrm{d}\mathbf{k} \mathrm{d}\mathbf{k}'' 
  A_0(\mathbf{k}) e^{-i\chi_1(\mathbf{k})} e^{-i\mathbf{k}\cdot \mathbf{x}'} 
  A_0(\mathbf{k}'') e^{i\chi_1(\mathbf{k}'')} e^{i\mathbf{k}''\cdot \mathbf{x}'}
  t(\mathbf{x}'+\mathbf{x}_0+\mathbf{x}_p).
\end{align}
So, with a structured probe,
\begin{equation} \label{eq:structured_t_M}
  t_M(\mathbf{x}_p+\mathbf{x}_0) = |a_1(-\mathbf{x}_p)|^2 \otimes t(\mathbf{x}_0+\mathbf{x}_p).
\end{equation}

\section{Contrast with more than two beams \label{sect:manybeams}}

We can get some better insight into the general case by considering the limit that the specimen transmission function does not vary over the scale of the probe size. In the limit that $A_m(\mathbf{k}) = 1$, the probes are infinitely small, and the interference pattern \eqref{eq:master} becomes
\begin{equation} \label{eq:DC_limit}
%  \left[A_m^{*}(\mathbf{k}) \otimes 
%  \left( T^{*}(\mathbf{k}) e^{-i\mathbf{k}\cdot\left(m \mathbf{x}_0 +\mathbf{x}_p\right)}\right)\right]
%  \left[A_n(\mathbf{k})\otimes 
%  \left( T(\mathbf{k}) e^{i\mathbf{k}\cdot\left(n \mathbf{x}_0 +\mathbf{x}_p\right)}\right)\right] 
  I_p(\mathbf{k}) \to  
  \sum_{\ell,m} c_{m}^* c_{\ell+m} 
  t^*(m\mathbf{x}_0+\mathbf{x}_p) t((\ell+m)\mathbf{x}_0+\mathbf{x}_p) 
  e^{-i\ell\mathbf{k}\cdot \mathbf{x}_0}.
\end{equation}
So, when the features of a specimen vary on a length scale much larger than the beam size (e.g. magnetic fields in an amorphous film), we see clearly that every plane wave in \eqref{eq:master} of order $\ell = n - m $ carries information from the interference of pairs of beams that have passed through the specimen at probe positions separated by a distance $\ell\,\mathbf{x}_0$. This limit matches the model we previously developed \cite{yasin_path-separated_2018}. 

% We noted in the main text that all pairs of beams separated by a distance $\ell \mathbf{x}_0$ contribute to a single plane wave in the recorded fringes. In the limit of probes much smaller than specimen features \ref{main-eq:DC_limit}, we concluded that
% \begin{equation} \label{eq:DC_limit}
%   I_p(\mathbf{k}) \to  
%   \sum_{\ell,m} c_{m}^* c_{\ell+m} 
%   t^*(m\mathbf{x}_0+\mathbf{x}_p) t((\ell+m)\mathbf{x}_0+\mathbf{x}_p) 
%   e^{-i\ell\mathbf{k}\cdot \mathbf{x}_0}.
% \end{equation}
We can rewrite this as
\begin{equation} \label{eq:DC_limit_reordered}
  I_p(\mathbf{k}) \to  
  \sum_{\ell} \left(\sum_{m} c_{m}^* c_{\ell+m} 
  t^*(m\mathbf{x}_0+\mathbf{x}_p) t((\ell+m)\mathbf{x}_0+\mathbf{x}_p) \right)
  e^{-i\ell\mathbf{k}\cdot \mathbf{x}_0}.
\end{equation}
If we have three dominant beams, i.e. $c_1,c_0,c_{-1} > 0$, we see that there is a term in $I_p(\mathbf{k})$,
\begin{equation} \label{eq:DC_limit_3beams}
  I_1(\mathbf{k}) = 
  \left(
  c_{0}^* c_{1}
  t^*(\mathbf{x}_p) t(\mathbf{x}_0+\mathbf{x}_p) +
  c_{-1}^* c_{0}
  t^*(-\mathbf{x}_0+\mathbf{x}_p) t(\mathbf{x}_p)
  \right)
  e^{-i\mathbf{k}\cdot \mathbf{x}_0}.
\end{equation}
These two terms can interfere. If the phase of the first term is $\pi$ different from the phase of the second term, there is zero contrast in the interference fringes with the lowest spatial frequency. We observed this in \cite{yasin_path-separated_2018}. The efficiency of imaging with more than two beams will therefore always be lower than with just two, as efficiency is linear with fringe contrast. However, there is still fringe contrast in the case just described, as the interference between $m=+1$ and $m=-1$ beams still produces fringes with twice the spatial frequency, as $\ell=2$ for this pair. This signal, as with the $\ell=1$ signal, allows for reconstruction of phase of the specimen with respect to vacuum. As adding an aperture to pass just two beams adds some complexity to the setup, it is worth considering in more detail the imaging capabilities with more than two beams.  

In the setup we previously demonstrated \cite{yasin_path-separated_2018}, with a phase diffraction grating that produces three dominant beams, it is possible to image with the $m=+1$ beam, with the $m=<1$ beams in vacuum for reference, and treat contributions from the much weaker $m>1$ beams as noise. However, we expect that more efficient imaging is possible with an iterative reconstruction process. For example, when the $m=2$ beam is on the specimen and the $m=1$ beam is in vacuum, the term $ c_{1}^* c_{2} t^*(\mathbf{x}_0+\mathbf{x}_p) t(2\mathbf{x}_0+\mathbf{x}_p)$ (where $t^*(\mathbf{x}_0+\mathbf{x}_p) = 1$) is redundant with the term $ c_{0}^* c_{1} t^*(\mathbf{x}_p') t(\mathbf{x}_0+\mathbf{x}_p')$ when $\mathbf{x}_p' = \mathbf{x}_0 + \mathbf{x}_p$, meaning the $m=1$ beam is on the specimen and the $m=0$ beam is in vacuum (i.e. $t^*(\mathbf{x}_p') = 1)$. Additionally, it is possible to reconstruct $t(2\mathbf{x}_0+\mathbf{x}_p)$ from the term $t^*(\mathbf{x}_0+\mathbf{x}_p) t(2\mathbf{x}_0+\mathbf{x}_p)$ when both $m=1$ and $m=2$ are on the specimen, as $t^*(\mathbf{x}_0+\mathbf{x}_p)$ can be reconstructed from the $t^*(\mathbf{x}_p) t(\mathbf{x}_0+\mathbf{x}_p)$ term. This additional information will be most easily recoverable with an iterative reconstruction that compares measured fringe patterns $I_p$ with expected fringe patterns at each probe position calculated from a model $t(\mathbf{x}_p)$.

\section{Scan artifact correction}

In an FEI Titan STEM like TEAM I, shifts of the probe on the specimen also lead to small shifts of the aperture $A_0(\mathbf{k})$ in the diffraction plane below the specimen, even with optimal alignment. As a shift of the aperture in the diffraction plane also causes a shift of the interference pattern in STEM holography, it is necessary to remove these shifts for an accurate reconstruction of the specimen transfer function. Fortunately, the position of the aperture in the diffraction plane is very nearly linear with the position of the beam on the specimen, so correction is straightforward.

In principle, determining the center of mass of the recorded electron counts and then shifting by offset at each probe position is possible. However, in our datasets, the lowest spatial frequency fringe spacing is on the order of ten pixels, so an error in shift correction of one pixel leads to an error in phase reconstruction on the order of 1 radian at that probe position. Accurate sub-pixel shifting would likely be slow. We instead calculated the transfer function at every probe position, including the artifactual shifts from the scan. These artifactual shifts appear as a linear phase in the transfer function. We fit this linear phase in a region of the scan where both probes are in vacuum, far from the sample, and then subtracted it everywhere. This method is fast, as it operates on 2D data rather than 4D data, and potentially more accurate than a center-of-mass measurement as it is insensitive to noise, cosmic rays, and other detector artifacts outside the aperture.

\begin{figure}[ht]
  \subfloat[]{\label{subfig:comparison:uncorr}
  \includegraphics[height=1.0in]{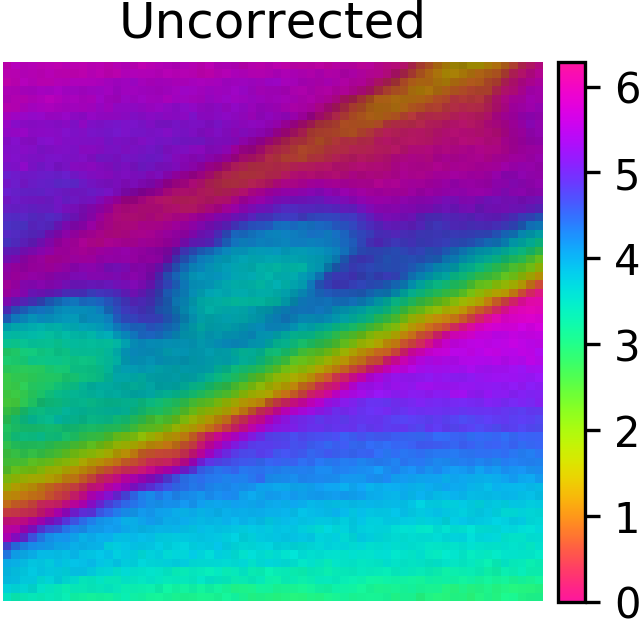} }
  \subfloat[]{\label{subfig:comparison:corr}
  \includegraphics[height=1.0in]{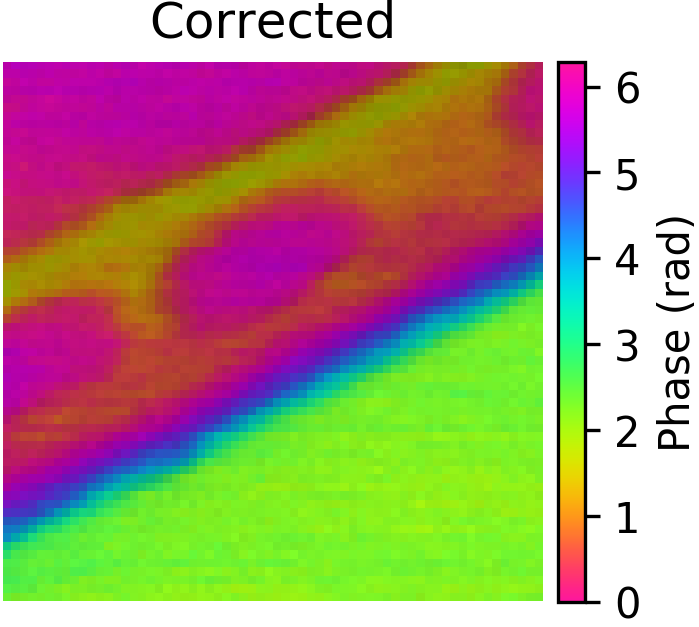} }
  \caption{Comparison of phase (color) without (a) and with (b) correction of the artifactual phase produced by the scan. (b) is an amplitude-phase version of the same data shown in \ref{fig:comparison}. \label{fig:correction}}
  % starting at the edge of the particle, as determined by the maximum gradient of the ADF. 
\end{figure}

\section{Dynamical Diffraction \label{sect:sim}}
We modeled the specimen as thin in describing it as a transfer function $t(\mathbf{x})$ indenpendent of the incident beam rather than a transfer matrix. Many specimens--in particular, crystals significantly thicker than an atomic monolayer--include dynamical diffraction effects that impact contrast in a variety of imaging modes. While a thorough investigation of dynamical diffraction effects on contrast in STEM holography is outside the scope of this manuscript, we can get a sense for how much these effects matter by examining the transfer function reconstructed from a simulated STEM holography experiment.
\begin{figure}
  \subfloat[]{\label{subfig:difference:phase}
  \includegraphics[height=1.0in]{2Beam_x0=15nm_phase.png}}
  \subfloat[]{\label{subfig:difference:model}
  \includegraphics[height=1.0in]{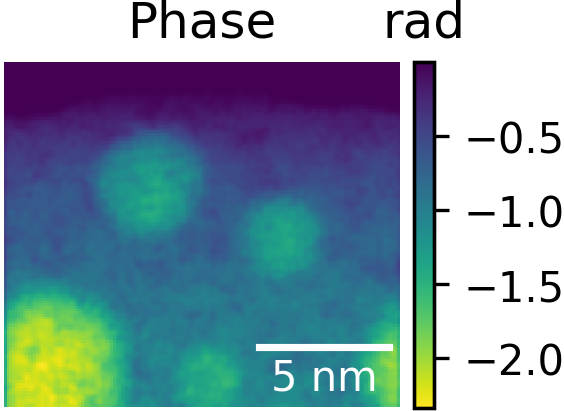}}
  \subfloat[]{\label{subfig:difference:diff}
  \includegraphics[height=1.0in]{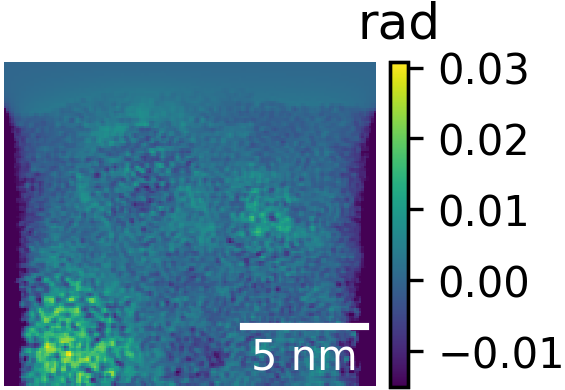}}
  \caption{Comparison of phase (a) reconstructed by STEM holography and (b) predicted from our model \eqref{eq:t_M} for a simulated STEM holography dataset. When we subtract (b) from (a), we can see that the differences (c) are largest on the thickest gold nanoparticle due to dynamical diffraction effects. The colorbar has been set to exclude the large difference at the edges due to boundary artifacts from the convolution used to generate (b). \label{fig:difference}}
\end{figure}

With a computer-generated specimen, we can compare the phase image we would expect based on our model and the projected potential of the specimen \eqref{eq:t_M}, and the phase image we reconstruct in a simulated STEM holography dataset. We see in Figure \ref{fig:difference} that the differences are small. The largest deviation is 1.5\% of the measured phase. The largest differences occur on the largest nanoparticle, where dynamical diffraction effects are strongest.
% If aberrations are not known and $\chi_c(\mathbf{k}) = 0$, the point spread function is instead
% \begin{equation}
%   \textrm{psf}(\mathbf{x}) =  a_a(-\mathbf{x}) a_a(\mathbf{x})
% \end{equation}

% So, if the aberration function $\chi(\mathbf{k})$ is known, aberrations can be corrected, up to the limit of the spatial coherence of the electron beam, with the appropriate choice of kernel. An aberration-free point spread function $\textrm{psf}(\mathbf{x}) = |a_a(\mathbf{x}|^2$ results.
% \begin{figure}
%   \subfloat[]{\label{subfig:lineplots:lacey}
%   \includegraphics[height=1in]{STEMh-HAADF_comparison_lacey_carbon.png} }
%   \subfloat[]{\label{subfig:lineplots:CdTe}
%   \includegraphics[height=1in]{STEMh-HAADF_comparison_CdTe.png} }\\
%   \caption{(a) Comparison of phase reconstructed by STEM holography (red) with HAADF signal (blue) on a lacey carbon edge. The noise in the phase is noticeably lower. (b) Comparison of phase (red) and HAADF (blue) on the edge of a semiconducting nanoparticle. Inset: zoomed-in plot of the vacuum region outside the edge of the particle. The phase is fit well by an exponential decay (green) starting at the edge of the particle, as determined by the maximum gradient of the HAADF. \label{fig:lineplots}}
% \end{figure}
% 

\end{widetext}
\bibliographystyle{apsrev4-1}
\bibliography{stemh_model}{}
\end{document}